\documentstyle[rotate,psfig,twocolumn]{mn}

\newcommand{\lapp}{\mbox{\raisebox{-0.3em}{$\stackrel{\textstyle <}{\sim}$}}}

\title[3C459: an asymmetric radio galaxy with a starburst]{3C459: A highly asymmetric radio galaxy with a starburst}

\author[P. Thomasson et al.]{P. Thomasson$^{1}$, D.J. Saikia$^{1,2}$ and T.W.B. Muxlow$^{1}$ \\
$^{1}$ The University of Manchester, Jodrell Bank Observatory, Macclesfield, Cheshire, SK11 9DL \\
$^{2}$ Tata Institute of Fundamental Research, National Centre for Radio
Astrophysics, P.B. No. 3, Ganeshkhind, Pune 411 007, India \\
}

\date{Received:}
\begin{document}
\maketitle

\begin{abstract}
Multifrequency radio observations of the radio galaxy 3C459 using MERLIN, VLA and the EVN, and an 
optical HST image using the F702W filter
are presented. The galaxy has a very asymmetric radio structure,
a high infrared luminosity and a young stellar population. 
The eastern component of the double-lobed structure is brighter, much closer to the nucleus
and is significantly less polarized than the western one. This is consistent with the jet on the eastern side
interacting with dense gas, which could be due to a merged companion or dense cloud of gas.
The HST image of the galaxy presented here exhibits filamentary structures, and is
compared with the MERLIN 5-GHz radio map.  EVN observations of the prominent central component,
which has a steep radio spectrum, show a strongly curved structure suggesing a bent or helical
radio jet. The radio structure of 3C459 is compared with other highly asymmetric,
Fanaroff-Riley II radio sources, which are also good candidates for studying jet-cloud interactions. 
Such sources are usually of small linear size and it is possible that the jets are 
interacting with clouds of infalling gas that fuel the radio source.
\end{abstract}

\begin{keywords}
galaxies: active - galaxies: jets - galaxies: nuclei - galaxies: individual: 3C459 - 
radio continuum: galaxies
\end{keywords}

\section{Introduction}
Although the majority of high-luminosity extragalactic radio sources selected at
a low frequency show symmetries in the brightness and location of the 
oppositely directed components, a small but significant fraction are highly asymmetric.
The asymmetries of the oppositely directed lobes of emission are 
important because they could provide useful insights into the environments
of sources and the interaction of jets with external clouds or galaxies. They also help in  
the understanding of the evolution of the individual components with time and 
provide tests of the orientation-based unified scheme for radio galaxies and quasars (Barthel 1989).
Using the well-known 3CR sample, McCarthy, van Breugel \& Kapahi (1991) have shown that
optical line-emitting gas tends to be brightest on the side on which the radio lobe is closer
to the nucleus, demonstrating the importance of environmental effects in the 
structural asymmetries of powerful radio sources. Polarization studies of these
lobes have shown that the nearer lobe also tends to depolarize more rapidly, possibly
due to interaction of the radio plasma with the line-emitting gas (cf. Pedelty et al. 1989,
Ishwara-Chandra et al. 1998). This suggests that the depolarization asymmetry of the lobes
is determined by an asymmetric environment as well as the effects of orientation
(Garrington et al. 1988; Laing 1988).

A class of sources that is of particular interest is the compact steep-spectrum sources
(CSSs), defined to be less than $\sim$20 kpc in size for a Universe with 
H$_0 =100$ km s$^{-1}$ Mpc$^{-1}$ and q$_0 = 0$ (cf. O'Dea 1998 for a review). These sources,
which are widely believed to be young radio sources,
tend to be more  asymmetric in both the brightness and location of the outer radio components
compared with the larger sources
(cf. Saikia et al. 2001). In a recent study of a sample of CSSs from the 3CR, S4, B2 
and B3 samples it has been shown that the CSSs exhibit large
brightness asymmetries with the flux density ratio for the opposing lobes being $>$5 for 
$\sim$25 per cent of the objects,
compared with only $\sim$5 per cent for the objects of larger size (Saikia et al. 2002).
The authors estimate the sizes of the clouds responsible for these asymmetries to
be $\sim$3 to 7 kpc, similar to those of dwarf galaxies, and speculate that such clouds
might be responsible for the infall of gas which fuels the radio source. Amongst 
larger-sized objects, one of the most asymmetric sources is B0500+630, the structure of which
appears to be inconsistent with the unified scheme and suggests that 
there is an intrinsic asymmetry in the oppositely-directed jets from the nucleus
(Saikia et al. 1996). The most extreme form of asymmetry is when the source is
completely one-sided with radio emission on only one side of the nucleus. 
Although most of these objects are core-dominated and their apparent asymmetry is
likely to be due to bulk relativisitic motion of the extended lobes of emission,
there do appear to be a number of weak-cored one-sided sources which are difficult
to reconcile with the simple relativistic beaming scenario (Saikia et al. 1990). 

\begin{figure*}
\hspace{0.0cm}
\begin{center}
\vbox{
\hspace{0.0cm}
\psfig{file=3c459hgeom-408-2.ps,width=5.0in,angle=-90}
\vskip -0.2in
\hskip 0.0 in \psfig{file=3c459-5000vla.ps,width=5.0in,angle=-90}
\hskip 0.5 in \psfig{file=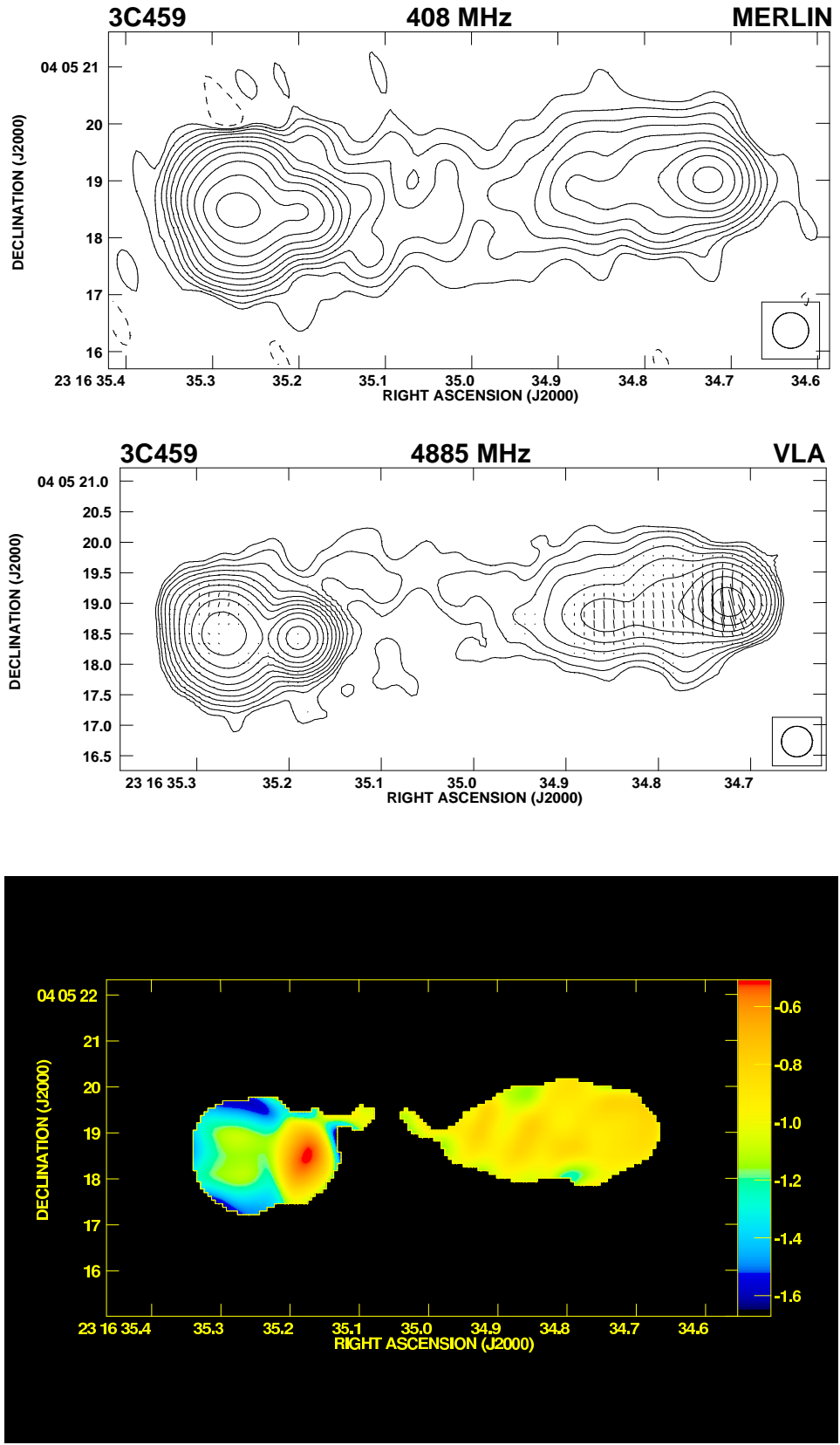,width=5.0in,angle=-90}
     }
\end{center}
\caption{Upper panel: The MERLIN image of 3C459 at 408 MHz with an angular resolution of 0.62
arcsec. Peak brightness: 4436 mJy/beam; Contour levels: 3$\times$($-$1, 1, 2, 4, 8 $\ldots$)
mJy/beam. Middle panel: The VLA image at 4885 MHz with a resolution of 0.50 arcsec. 
Peak brightness: 382 mJy/beam; Contour levels: 0.25$\times$($-$1, 1, 2, 4, 8 $\ldots$) 
mJy/beam.  Polarization: 1 arcsec = 25 mJy/beam. 
Lower panel: The spectral index distribution between 408 and 4885 MHz made
by convolving the VLA image to a resolution of 0.62 arcsec.
}
\end{figure*}

As part of a study of highly asymmetric radio sources, the radio galaxy 3C459 has been
observed extensively at radio and optical wavelengths. The results of these observations 
are presented in this paper. 3C459 is identified with a 17.55 
V magnitude,  N-galaxy at a redshift of 0.2199 (Spinrad et al. 1985; Eracleous \&
Halpern 1994), so that 1 arcsec corresponds to 2.39 kpc.
Its optical/UV spectrum is dominated by the light of
a young stellar population, with the Balmer break and higher Balmer absorption lines
being detected (cf. Miller 1981; Tadhunter et al. 2002). 
Although Yee \& Oke (1978) reported both H$\alpha$ and H$\beta$ to exhibit a broad, low
profile, no clear evidence of broad permitted lines were found by Eracleous \& Halpern (1994),
and more recently by Tadhunter et al. (2002). Both of these sets of authors as well as 
Heckman et al. (1994) have clasified 3C459 as a narrow-line radio galaxy. 
Its far-infrared luminosity
at 60$\mu$m is unusually high, being about 10 times brighter than other radio sources
from the 2-Jy sample at comparable redshifts (cf. Tadhunter et al. 2002). Tadhunter et al.
also note a possible relationship between optical/UV starburst activity and far-infrared excess, 
and suggest that the high infrared luminosity is due to dust heating by
the starburst. 
The source has been detected with ISO by Fanti et al. (2000); and HST F702W and WFPC2 V-band
images have been presented by de Koff et al. (1996) and Farrah et al. (2001) respectively.
The radio galaxy has also been detected in HI absorption with the VLA
as well as by the WSRT (Morganti et al. 2001). The FWHM of the absorption seems to be
quite broad with a width of $\sim$400 km s$^{-1}$. The integrated HI column density 
of $\sim$2.7$\times$10$^{18} T_{\rm spin}$ cm$^{-2}$ is similar to that found for other
radio galaxies.
It is interesting to note that the gaint radio galaxy 3C236 also shows evidence of star 
formation, and HI absorption against a lobe of the inner radio source (Conway \& Schilizzi 2000;
Schilizzi et al. 2001; O'Dea et al. 2001).
 
The radio structure of this galaxy has been studied by Ulvestad (1985, hereinafter
referred to as U85) using the VLA.
His results have shown that the radio structure of the source comprises a core and two
extended lobes, the eastern one being a factor of $\sim$5 closer to the core than the
western lobe and the whole source extending to approximately 8.2 arcsec. 
This corresponds to a linear size of 19.5 kpc, which is similar to other compact 
steep-spectrum objects. At the VLA resolution of $\sim$0.4 arcsec at $\lambda$6 cm, 
the eastern lobe, though significantly resolved, appears to have a smooth structure 
with no discernible small-scale features. The western lobe, extending half-way back
to the core has a tail with two peaks of emission.  The higher-resolution
$\lambda$2 cm image of U85 shows the source to have an edge-brightened,
FRII structure, consistent with its radio luminosity of 2.1$\times$10$^{25}$ W Hz$^{-1}$
sr$^{-1}$ at 1400 MHz. It is worth noting that the central component, which contributes $\sim$30 
per cent of the total flux density at 5 GHz, has a steep radio spectrum with a spectral index,
$\alpha$ (S$\propto\nu^{\alpha}$), of $-$0.78$\pm$0.15 between $\lambda$6 and 2 cm
(U85). 3C459 also exhibits
a high  degree of polarization asymmetry between the two lobes (Davis, Stannard \& 
Conway 1983; U85; Morganti et al. 1999). It has an integrated rotation measure
of $-$6$\pm$1 rad m$^{-2}$ with an intrinsic position angle (PA) of 7$\pm$3$^{\circ}$
(Simard-Normandin, Kronberg \& Button 1981). 

The observations and analyses of the data are described in Section 2, and the 
results are presented in Section 3. Possible explanations for the structure are
discussed in Section 4, while the conclusions are summarised in Section 5.

\begin{table}
\caption{The observing log}
\vspace*{1ex}
\begin{tabular}{l l l l }
Telescope &  Obs.  & Antennas                    & Obs.        \\
          &  Freq. &                             & Date        \\
          &  MHz   &                             &             \\
\\[0.5mm]
MERLIN    &  408   &  C1,De,Kn,Lo,Ta,Wa          & 1987 Oct 29 \\
MERLIN    & 1420   &  C32,Da,De,Kn,Lo,           & 1993 Nov 07 \\
          &        &                  Mk2,Ta,Wa  &             \\
MERLIN    & 1658   &  C32,Da,De,Kn,Lo,           & 1993 Nov 07 \\
          &        &                  Mk2,Ta,Wa  &             \\
MERLIN    & 4546   &  C32,Da,De,Kn,Mk2,Ta        & 1995 Jul 02 \\
MERLIN    & 4866   &  C32,Da,De,Kn,Mk2,Ta        & 1995 Jul 02 \\
MERLIN    & 4993   &  C32,Da,De,Kn,Mk2,Ta        & 1992 Jul 19 \\
MERLIN    & 5186   &  C32,Da,De,Kn,Mk2,Ta        & 1995 Jul 02 \\
EVN-VLBI  & 4987   &  Ef,Me,Mk2,No,On,Wk         & 1995 May 24 \\
\end{tabular}

\vspace*{1ex}
{\bf Antennas}: C1 One antenna of the Cambridge one-mile telescope, 
C32 Cambridge 32m, 
Da Darnhall, De Defford, Ef Effelsberg, Kn Knockin, Lo Lovell, Me Medicina, 
Mk2 Jodrell Bank Mk2, No Noto, On Onsala, Ta Tabley, Wa Wardle, Wk Westerbork 
\end{table}

\section{Observations and analyses}

The MERLIN observations presented in this paper have been made over a period of time from 1987 October 
to 1995 May.  They cover a frequency range from 408 MHz to $\sim$5 GHz, with resulting image 
resolutions ranging from 620 mas at 408 MHz to $\sim$70 mas at 5 GHz.   The size of 3C459 and its 
low declination ($\sim$4 degrees) make it a difficult source to map with MERLIN alone at 5 GHz.   
Thus, shorter spacing calibrated VLA A-array data, kindly provided by Ulvestad, have been combined 
with the MERLIN data to provide additional constraints in the `mapping' process.   Even so, it is
still not possible to fully remove the north-south sidelobes associated with the strong core. 
An improved 
`VLA only' 5 GHz image with a resolution comparable to that of the MERLIN 408 MHz image has also 
resulted from a reprocessing of the VLA data.   Since 3C459 is known not to be variable, 
there is no difficulty in combining data or comparing the resulting 
images from the different frequencies or instruments over the period of the observations.   5 GHz 
VLBI observations have also been made with the EVN in May 1995, yielding an image of the core 
region of the source with a resolution of 6 mas.   The dates of the MERLIN and VLBI 
observations and  
the antennas used are given in Table 1.   
The flux calibration sources are 3C48 and 3C286, their flux densities being calculated from the 
values given in  the VLA calibrator source catalogue, which is 
based on the work of Baars et al. (1977).   
The point source baseline calibrator fluxes have been determined from a comparison of their MERLIN 
short spacing amplitudes with those of 3C48 or 3C286.   3C286 is assumed to have a position angle 
for its linear polarization of 32.4$^\circ$. The baseline calibrators used were B1345+125 at 408 MHz and
B0552+398 and OQ208 at the higher frequencies.  The phase reference source used for all the MERLIN 
and EVN observations except that at 408 MHz was B2318+049.   Its position, obtained from the 
IERS93 list of VLBI calibrators, is J2000 23$^h$ 20$^m$ 44.$^s$85661  $+$05$^\circ$13$^\prime$ 
49.$^{\prime\prime}$49529.   An optical continuum WFPC2 image ( Filter F702W ) has been retrieved 
from the HST archive to provide a high resolution comparison between the radio and optical emission. 

\subsection{MERLIN 408 MHz Observation}
     The 408 MHz observation of 3C459 was made on 1987 October 29, i.e. during the period of 
upgrade of the original MERLIN (Thomasson 1986) when, with microwave links to Cambridge newly installed, it was 
possible to use one of the 18m  `one-mile' synthesis telescopes at Cambridge with the original 
network and correlator prior to the completion of the new 32m antenna.   As the original correlator 
could only be used to correlate data from 6 telescopes at that time because of a fault, the 
Lovell and Wardle telescopes, with their greater sensitivity at 408 MHz, were included in the network 
in preference to the somewhat similarly located Mk2 and Darnhall telescopes.   The observing bandwidth was 
5 MHz centred at 408 MHz and the polarization was limited to a single one (LL)  by the frequency-modulated 
microwave link.   Phase-referenced observations were not possible using the Cambridge 18m telescope 
and so 3C459 was simply followed for approximately the period of time that it was above the horizon ($\sim$ 11 hrs.).   
In processing the data using the Jodrell Bank OLAF and NRAO AIPS packages, small corrections were 
made for the sky background temperatures using the 408 MHz all-sky survey of Haslam et al. (1982).   
The final 408 MHz image is shown in Figure 1.   Since 
the observations were made without reference to a source of known position, the absolute position could 
not be determined from these observations.   However, the positions shown in Figure 1 have been established 
from a comparison of the mean positions of the core and eastern lobes with that of the VLA image at 5 GHz 
at the same resolution.

\subsection{MERLIN L-Band Observations}
As indicated above, the {\it uv} coverage of MERLIN for a single frequency full-track observation of 3C459 is rather poor with quite large gaps. In an attempt to improve this at L-Band frequencies, 3C459 was observed with MERLIN at two L-Band frequencies, 1420 MHz and 1658 MHz. The observing frequency was switched between the two frequencies every 5.5 minutes with a total cycle time, including the time on the phase reference source, of 11 minutes. The source and its phase reference were observed for a total of $\sim$10 hours 
in all four polarisations (LL, LR, RL and RR) with all the MERLIN telescopes included in the array. This resulted in a total of $\sim$3 hours actual 
on-source time at each frequency.   Lovell telescope drive restrictions and the very slow maximum speed 
of the Wardle telescope meant that these two telescopes were only moved to the phase reference source 
approximately every half hour.   The changes in phase of the signals from these two telescopes during 
this half hour period were effectively tracked by comparison with the phase changes of the `nearby' 
Mk2 and Darnhall telescopes respectively.   All the data were processed using the Jodrell Bank 
D-programmes (for the initial editing and calibration) and the MERLIN pipeline, which uses AIPS tasks, 
to produce initial images of 3C459 at the two frequencies.   The 1420 MHz data amplitudes were then scaled to those of the 1658 MHz data with appropriate scaling factors based upon the relative amplitudes of the source at the same 
resolution at the two frequencies. Final total power and polarization images of 3C459 at a nominal frequency of 1658 MHz were 
produced (Figure 2) using the self-calibration and imaging routines in AIPS.

\subsection{C-Band Observations}
\subsubsection{VLA A-array Observations}
The calibrated VLA A-array data at 4885 MHz, kindly made available by Ulvestad, resulted from 
an observation on 1983 September 8.  These data were further edited and a new 0.5 arcsecond 
resolution image produced using the self-calibration and IMAGR tasks in AIPS.   Our final image, 
which shows rather more of the extended emission than that in the original map of Ulvestad, 
is shown in Figure 1.

\subsubsection{MERLIN Observations}
     3C459 was observed with MERLIN at 4993 MHz on 1992 July 19 and, to improve the UV coverage, 
at 4546 MHz, 4866 MHz and 5186 MHz on 1995 July 2.   The telescopes configured in the array 
are given in Table 1.   A somewhat similar cycle to that at L-Band was 
used for switching round the three frequencies and between 3C459 and its phase reference source 
on July 2.   This resulted in $\sim$6 minutes on source integration at each frequency per half 
hour of elapsed time.   For the full observation of $\sim$11 hours, the integration time on 3C459 
at each frequency was $\sim$2.75 hours.   The on-source integration time for the 4993 MHz observation 
was $\sim$7.75 hours.   As for the L-Band data processing, images of 3C459 were made at the same 
resolution at each of the four frequencies using the D-programmes for editing and calibration and 
the MERLIN pipeline and the AIPS self-calibration and IMAGR tasks for the imaging.   The data 
at the four different frequencies were then combined together correcting for source spectral index
with appropriate scaling factors based 
upon the relative ampltudes of the unresolved core of the source in the preliminary images.
Finally, the VLA data were combined with the MERLIN 
data to yield the final maps at a resolution of 0.07 arcseconds shown in Figure 3.

\subsubsection{EVN Observations}
     A MERLIN$+$EVN observation of 3C459 was made at 4988 MHz on 1995 May 24.   The telescopes included 
in the EVN are given in Table 1.   Only one polarisation was recorded, LL, and the bandwidth was 28 MHz.   
The Cambridge telescope was originally included in the system, but unfortunately this failed and thus 
left a significant gap between the MERLIN and EVN UV coverage.   Consequently, only an EVN map of the 
core region has been produced.   The data were correlated in Bonn and further processed using the 
standard AIPS VLBI calibration, fringe fitting and imaging routines.   The calibration sources used were 
0552+398 and OQ208, though the flux scale was established from receiver noise measurements.   After 
initial calibration, the phase reference source, 2318+049, was imaged using the standard AIPS  
self-calibration and IMAGR routines.   It was found to be only slightly resolved.   The data for 
3C459 were corrected for the telescope  amplitude and phase variations determined from the 
phase-reference source and an image of the core region of 3C459 at a resolution of 6 mas was produced 
(See Figure 4).   The positional accuracy of the image is limited by inaccuracies in the Bonn 
correlator model and the position of 2318+049. A small improvement in 
the 3C459 image was obtained as a result of a further self-calibration cycle allowing the telescope 
phases to vary, which indicated small errors.   However, it should be noted that the absolute position 
of the image is not of importance for this paper.

\subsection{HST Observations}

     A WFPC2 continuum image, obtained on 1994 October 11 using the F702W filter as part of a programme 
for W. B. Sparks, was extracted from the HST archive.   
The bandpass of the F702W filter includes H$\alpha$ and possibly [OIII] emission (de Vries et al. 1999).
The image was further processed using 
standard routines in IRAF and a FITS file produced.   This was read into AIPS and the image 
re-gridded to be the same as the MERLIN + VLBI C-Band image.    The image is shown in Figure 5. 
 
%
%
\begin{table*}
\caption{Observational parameters and observed properties} 

\vspace*{1ex}
\begin{tabular}{l r c c          l lll rrl rr r}
\multicolumn{4}{c}{Observational parameters} & \multicolumn{10}{c}{Observed properties} \\
Telescope & Freq. & Resn. & $\sigma$    &  Cmp   &  \multicolumn{3}{c}{RA(J2000)} & \multicolumn{3}{c}{Dec(J2000)} & S$_p$ & S$_t$ & m     \\
  & MHz & $^{\prime\prime}$ & mJy/b     &        & h & m & s & $^{\circ}$ &  $^{\prime}$ &  $^{\prime\prime}$ & mJy/b &  mJy & \%          \\
(1)       &  (2)  &  (3)    &  (4)      &   (5)  & \multicolumn{3}{c}{(6)} & \multicolumn{3}{c}{(7)}          &   (8) &  (9) &  (10)       \\
\\[1mm] 

MERLIN    & 408   &  0.620  & l.05      &  W       & 23 & 16 & 34.73   & 04 & 05 & 19.05 &   1019 &  2650 &      \\
          &       &         &           &  C$^g$   &    &    & 35.19   &    &    & 18.43 &   1564 &  2220 &      \\
          &       &         &           &  E$^g$   &    &    & 35.27   &    &    & 18.49 &   4251 & 10490 &      \\

VLA       & 4885  &  0.500  & 0.08      &  W       & 23 & 16 & 34.73   & 04 & 05 & 18.97 &     96 &   265 & 17.8      \\
          &       &         &           &  C$^g$   &    &    & 35.19   &    &    & 18.43 &    381 &   392 & $<$0.2    \\
          &       &         &           &  E$^g$   &    &    & 35.27   &    &    & 18.50 &    203 &   590 & $\sim$0.9 \\

MERLIN    & 1658  &  0.225  & 0.22      &  W       & 23 & 16 & 34.73   & 04 & 05 & 18.84 &    126 &   723 & 13.0      \\
          &       &         &           &  C$^g$   &    &    & 35.20   &    &    & 18.31 &    775 &   873 & $\sim$0.4 \\
          &       &         &           &  E       &    &    & 35.29   &    &    & 18.30 &    247 &  2469 & 1.2       \\

MERLIN+   & 4866  &  0.070  & 0.07      &  W       & 23 & 16 & 34.73   & 04 & 05 & 18.81 &    9.4 &   300 & 20.0      \\
VLA       &       &         &           &  C1$^g$  &    &    & 35.19   &    &    & 18.31 &    390 &   413 & $\sim$0.3 \\
          &       &         &           &  C2$^g$  &    &    & 35.21   &    &    & 18.24 &    6.8 &    19 & $\lapp$4.0   \\
          &       &         &           &  E       &    &    & 35.29   &    &    & 18.33 &     15 &   674 & $\sim$2.6       \\
 
EVN-VLBI  & 4987  &  0.006  & 0.18      &  C1A$^g$ & 23 & 16 & 35.1942 & 04 & 05 & 18.446 &     25 &    55 &           \\
          &       &         &           &  C1B$^g$ &    &    & 35.1944 &    &    & 18.438 &     75 &   127 &           \\
          &       &         &           &  C1C$^g$ &    &    & 35.1949 &    &    & 18.435 &     84 &   132 &           \\
          &       &         &           &  C1D$^g$ &    &    & 35.1960 &    &    & 18.437 &     10 &    23 &           \\

\end{tabular}
\end{table*}
%

\begin{figure*}
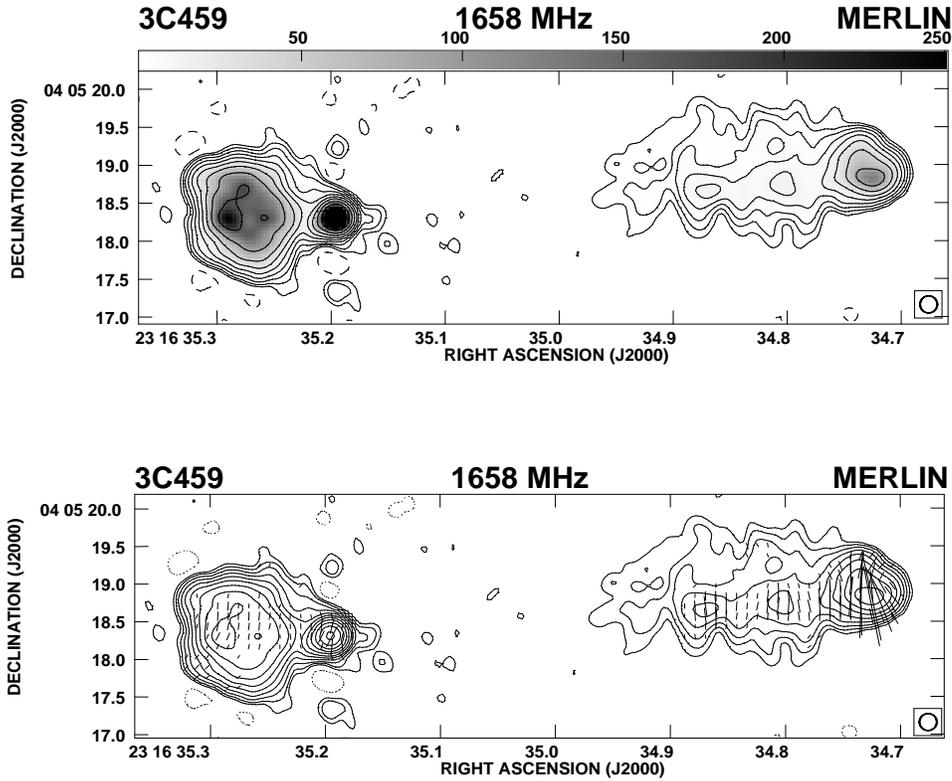

\begin{center}
\vbox{
\hspace{0.0cm}
\psfig{file=3c459-lband-grey.ps,width=5.0in,angle=-90}
\hspace{1.8cm}
\psfig{file=3c459-lband-pol.ps,width=5.0in,angle=-90}
}
\end{center}
\caption{The MERLIN images of 3C459 at 1658 MHz with an angular resolution of
225 mas. The upper panel shows the
total-intensity image while the lower one shows the polarization E-vectors
superimposed on the total intensity contours. Peak brightness = 769 mJy/beam;
Contours = 0.66$\times$($-$1, 1, 2, 4, 8 $\ldots$) mJy/beam. Polarization:
1 arcsec = 11.1 mJy/beam.
}
\end{figure*}

\section{Discussion}

Some of the observational parameters and observed values from the radio 
observations are summarised in Table 2. This table is arranged as follows.
Column 1: telescope used for the observations; column 2: observing frequency in MHz;
column 3: the angular resolution in arcsec;
column 4: the rms noise in the image in units of mJy/beam; column 5: component
designation, with a superscript $g$ indicating that the values in columns 6 to 9 have
been estimated by a fitting a two-dimensional Gaussian; columns 6 and 7: the right 
ascension and declination of the peak of emission 
of the component in J2000 co-ordinates; columns 8 and 9: the peak and total flux density
of the component in units of mJy/beam and mJy respectively; column 9: the degree of 
polarization of the component estimated by integrating the polarized- and total-intensity
images over identical boxes around each component for the extended lobes. The values for
the central compact components are at the pixels of maximum brightness.

\begin{figure*}
\begin{center}
\vbox{
\hspace{0.0cm}
\psfig{file=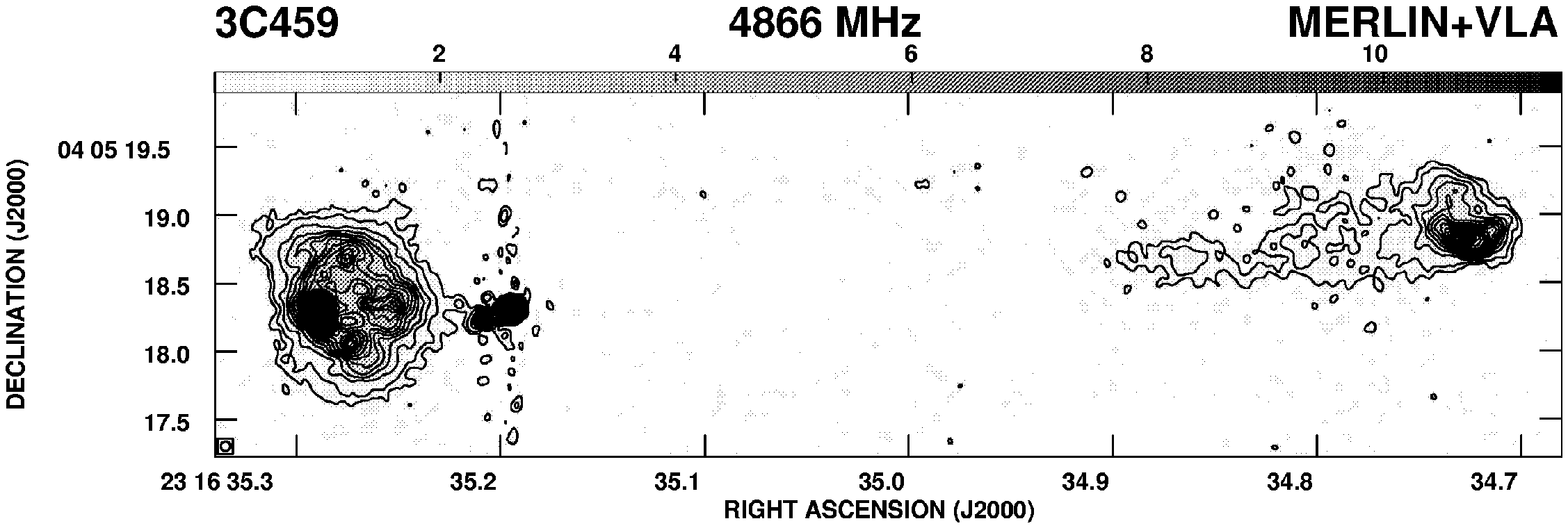,width=7.0in,angle=-90}
\vspace{0.3cm}

         \hbox{
               \psfig{file=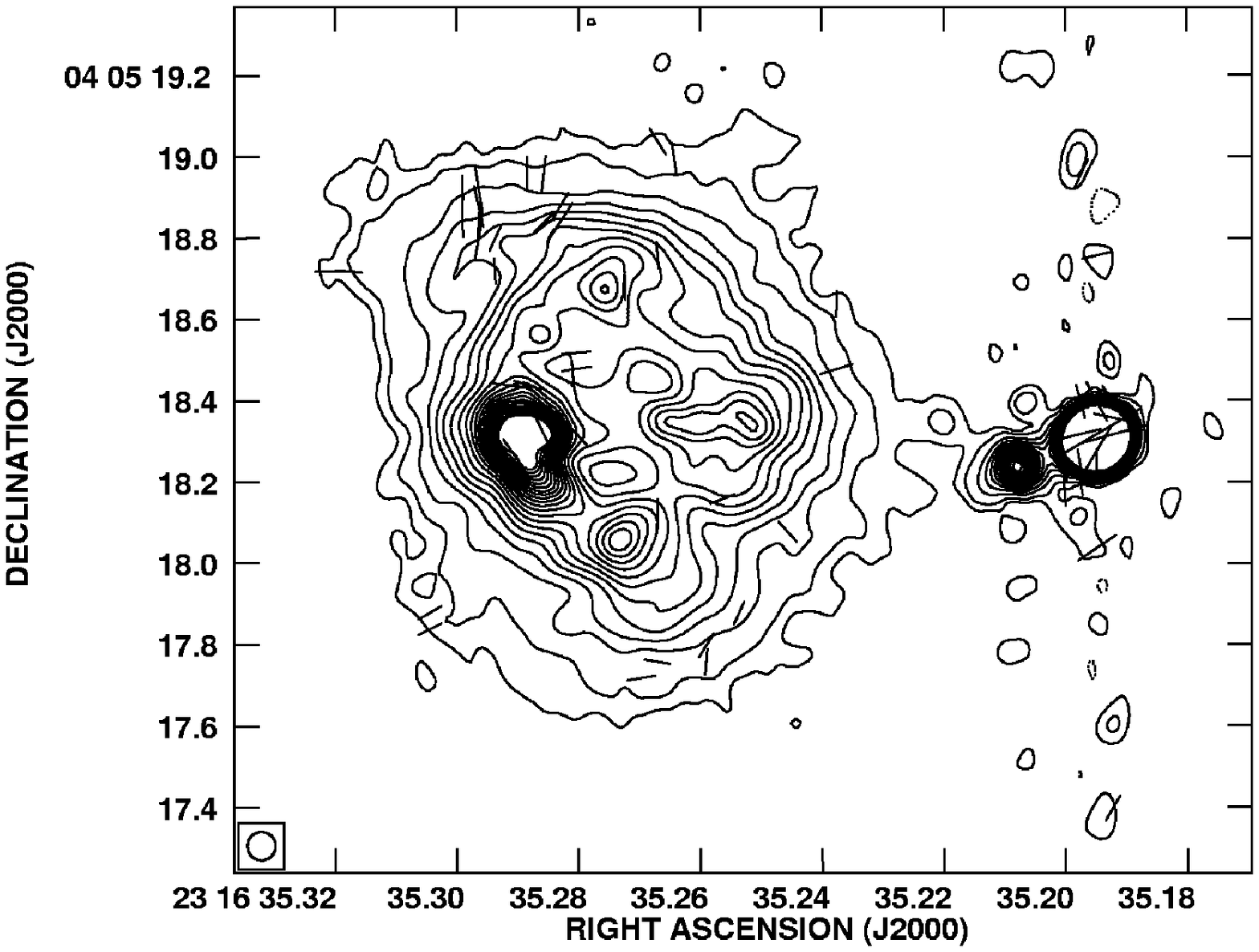,width=3.0in,angle=-90}
               \psfig{file=3c459w-cband-pol.ps,width=4.0in,angle=-90}
              }
}
\end{center}
\caption{The MERLIN images of 3C459 at 4866 MHz with an angular resolution of 70 mas. The 
total-intensity image is shown in the upper panel, while the lower panel shows the components 
with the polarization E-vectors superimposed on the total intensity contours.
Peak brightness: 391 mJy/beam; Contours: 0.29$\times$($-$1, 1, 2, 4, 6, 8, 10
$\ldots$ 36, 38, 40) mJy/beam. Polarization: 1 arcsec = 5 mJy/beam.
}
\end{figure*}

\subsection{The overall radio structure}

The MERLIN image of 3C459 at 408 MHz with an angular resolution of 0.62 arcsec 
is presented in Figure 1, which shows the well-known
double-lobed structure and the central component. In addition, there is a 
continuous bridge of emission which is
seen extending from the western lobe to the eastern one. The VLA A-array image with the
polarization vectors superimposed on the total intensity contours, and a spectral-index map
between 408 and 4866 MHz,
made by convolving the VLA image to a resolution of 0.62 arcsec, are also shown in Figure 1.
The bridge of emission is now partially seen in the VLA image as well. 
The western lobe is more strongly polarized with the peak in the hotspot being 13.3\%
polarized. The tail of emission from the western hotspot is the most strongly polarized 
feature with typical values ranging from $\sim$25 to 30\%. The low rotation measure (U85), 
suggests that the magnetic field lines are along this feature, similar to those seen in the jets of
FRII radio sources. There is no significant polarization detected from the central component,
the percentage polarization,$m$, at the total-intensity peak being $<$0.1\%. The eastern lobe is
weakly polarized with $m<$0.2\% at the total-intensity peak, but there are regions of emission
which are up to $\sim$5 and 2\% polarized to the north and east respectively of the peak.

The spectral index of the western lobe is $\sim-$0.9 near the peak of emission and varies
between approximately $-$0.8 and $-$0.9 along the ridge of emission extending eastwards. 
The central component has a spectral index of $-$0.6 near its peak, while the eastern lobe
has the steepest spectrum with $\alpha\sim-$1.2 in the central region and steepening to 
$\sim-$1.6 in the northern and southern ends of the lobe. The marginally flatter regions
with $\alpha\sim-$1.1 towards the north and south of the total-intensity peak in the eastern
lobe are close to the regions of higher brightness seen in the higher-resolution images.
However, the prominent hotspot seen in the higher-resolution image has $\alpha\sim-$1.2.
U85 also find the eastern component to have a steeper spectrum than the western one, 
the average values of $\alpha$ being $-$1.4 and $-$0.95 respectively. Our values are
somewhat smaller, suggesting a steepening of the spectrum towards higher frequencies.
However, there is no evidence of a spectral index as steep as $-$1.65 between the central
component and the eastern lobe as noted by U85.

The MERLIN image at 1658 MHz which has an angular resolution of 0.225 arcsec
(Figure 2) reveals greater details of the structure. A two-dimensional Gaussian fit to
the central component shows it to be extended along a PA  of 100$^\circ$, while the extension
towards the south-east is along a PA of $\sim$110$^\circ$. The western component contains
a curved high-brightness region of emission at the outer edge with the field lines 
also appearing to follow the bend. In addition there are two peaks along
the ridge of emission pointing towards the central component. The polarization vectors are
consistent with the low integrated value of RM and also the low value of RM estimated
for this lobe by U85.  

One of the striking features of the image at 1658 MHz is the shell-like structure of
the eastern component where the central region of the lobe is of lower surface 
brightness than the surrounding features. The 4866 MHz image with an angular resolution
of 70 mas, which has been made by combining the MERLIN and VLA data (Figure 3),
shows greater details of the eastern lobe with several distinct components surrounding
the region of lower surface brightness. 
The peak of emission on the western side of the eastern lobe
at RA 23$^h$ 16$^m$ 35.$^s$25, Dec 04$^\circ$ 05$^\prime$ 18.$^{\prime\prime}$34 
has a narrow-elongated ridge of emission which 
would meet Bridle \& Perley's (1984) criterion of being called a jet. This feature,
which we identify as part of a radio jet, points directly at the eastern
hot-spot which is the brightest component in the eastern lobe. This jet also appears to
be present outside of the shell and to be connected to the core region. The two other 
peaks of emission in the lobe north and south of the hotspot and the jet are likely to be
secondary hot-spots caused by outflow from the prominent one.
The image of the western lobe at 4866 MHz shows the hotspot to have a C-shaped structure,
with the  field lines following the curvature. The field lines are possibly sheared to
follow the direction of fluid flows, suggesting that the hotspot structure is caused by 
outflows from the point of impact of the jet from the nucleus. 

Hydrodynamic simulations of light, large-scale
jets in a decreasing density profile, 
which have also been examined by Carvalho \& O'Dea (2002),
show that the jet bow shock undergoes two phases,
firstly a nearly spherical one and secondly the well-known cigar-shaped one
(Krause 2002; Krause \& Camenzind 2002).  The shell-like structure of the eastern lobe
is suggestive of 
the first phase of the development of the bow-shock. In this scenario, the eastern jet
has not yet entered the cigar phase and deposits its radio-emitting plasma in a bigger
part of the bubble, almost filling the region within the bounds of the bow shock. On 
the other hand, the western jet appears to be in the cigar phase, and should therefore
have a fairly regular backflow around it, which flows back into the central parts diffusing
and mixing with the shocked external gas. 

In the highest resolution image at 5 GHz, the peak in the hotspot in the western lobe 
is $\sim$18\% polarized, while the corresponding feature in the eastern lobe is 
$<$2\% polarized. Besides the asymmetry in the location and brightness of the outer
components, another striking feature of this source is the polarization asymmetry. 
Since the images with radio polarization information are of very different resolutions, 
it is not possible to derive reliable values of depolarization. However, it is clear that 
the western lobe is only
slightly depolarized between 5 and 1.7 GHz, while the eastern lobe is strongly 
depolarized by $\sim$5 GHz. 
3C459 is consistent with the Liu-Pooley relationship (Liu \& Pooley 1991), which shows that
the radio lobe with a flatter  radio spectrum is less depolarized. This relationship is
significantly stronger for smaller sources, but is similar for both radio galaxies and quasars
suggesting that in addition to Doppler effects there are intrinsic differences between the
lobes on opposite sides (cf. Ishwara-Chandra et al. 2001).  
Assuming that the extension of the core towards the south-east
and the jet-like feature within the lobe defines the jet direction to be towards the 
east, 3C459 is not consistent with the Laing-Garrington effect (Laing 1988; Garrington et al. 
1988). This is not surprising if the external environment is very asymmetric with the
eastern jet interacting with dense gas which slows down the jet and also
depolarizes the radio emission. It is also to be noted that in the 1658-MHz image there is
a possible indication of a jet-like structure pointing towards the western lobe from the core.
 
\begin{figure}
\begin{center}
\vbox{
\hspace{0.0cm}
\psfig{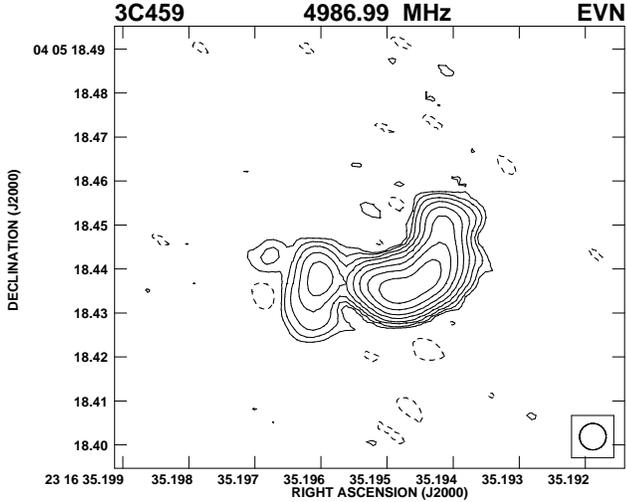}
}
\end{center}
\caption{The EVN-VLBI image of 3C459 at 4987 MHz with an angular resolution of 6 mas.
Peak brightness: 95 mJy/beam; Contours: 0.44$\times$($-$1, 1, 2, 4, 8 $\ldots$) mJy/beam.
}
\end{figure}

\subsection{The central component}

The central component is
clearly resolved into two distinct components with a separation of $\sim$200 mas along a
PA of 110$^\circ$. The dominant component (C1) is more compact with a deconvolved size
of 23$\times$9 mas along a PA of 117$^\circ$ compared with 130$\times$55 mas along a
PA of 108$^\circ$ for the weaker one (C2). The peak brightness of C1 is also higher than that
of C2 by $\sim$55, and perhaps contains the true nucleus or radio core of this galaxy. 
As noted earlier, the overall spectrum of the central component
is steep between 400 and 5000 MHz with a spectral index of $\sim-$0.6. Since the
stronger component, namely C1, dominates the flux density of the entire central 
component seen in the lower-resolution images, C1 must have a steep spectral index 
although multifrequency, high-resolution data are not available to determine the spectral
indices of C1 and C2 separately. 

The EVN image of the dominant central component, C1, (Figure 4) shows a rather complex 
structure with at least four components along a strongly curved ridge of emission. 
The components are within a factor of 5 in brightness, and it is not clear which, if any
of these components, represents the true nucleus of the galaxy. If the jet is one-sided,
as in most high-luminosity radio sources, then the northernmost component could be the
true nucleus of the galaxy. Multi-frequency data with similar resolution would be required
to confirm this possibility. However, if this is the case, the jet swings from an initial
PA of 165$^\circ$ to $\sim$120$^\circ$, defined by the two prominent peaks, and later to
$\sim$80$^\circ$. Considering that C2 is at PA of 110$^\circ$ and then goes northwards,
the jet appears to have a helical structure. No emission was detected in the VLBI image
at the position of C2.

\subsection{The HST image and comparison with the radio image}

The HST image of 3C459 using the F702W filter is shown in Figure 5. Its position 
has been determined by aligning the dominant nucleus of this N-galaxy with the core
component, C1, seen in the MERLIN+VLA image (Figure 3). In addition to the nucleus,
there is a prominent secondary peak east of the nucleus at 
RA 23$^h$ 16$^m$ 35.$^s$233, Dec 04$^\circ$ 05$^\prime$ 18.$^{\prime\prime}$22,
and a filamentary structure extending from near this peak towards the east. 
A less prominent filamentary structure is also seen extending north at
$\sim$RA 23$^h$ 16$^m$ 35.$^s$22, Dec 04$^\circ$ 05$^\prime$ 18.$^{\prime\prime}$8, 
and there are also other weaker peaks of emission in the HST image.
The secondary peak is separated from the nucleus by 575 mas along
a PA of 99$^\circ$. Although this lies well beyond the weaker 
radio component in the nucleus, C2, and does not correspond to any obvious feature in the
radio image, it could affect the path of the jet (Figure 6). The eastern filament,
skirts the lower end of the eastern radio lobe.

A ground-based V-band image of 3C459 with the CTIO 4m telescope shows filamentary  or
`fanlike protrusion' extending $\sim$8$^{\prime\prime}$ to the east and a similar
but more knotty feature towards the south (Heckman et al. 1986). Long-slit spectroscopic
observations along and perpendicular to the radio source axis and passing through the
east and south fans show the emission line gas to be fairly compact. This led Heckman
et al. to suggest that the  fans are continuum-emitting structures. There are no other
galaxies nearby with which it could be interacting, suggesting that if these morphological 
peculiarities are indeed of tidal origin 3C459 could be a case of a merger which has
nearly reached completion. The secondary peak of emission seen in the HST image could
be the  merging galaxy buried in the debris. The possibility that gravitational 
interactions between galaxies might trigger nuclear activity has had a long 
lineage (e.g. Baade \& Minkowski 1954; Toomre \& Toomre 1972; Quinn 1984; Hernquist \& Mihos 1995),
and 3C459 with its optical morphology, high infrared luminosity, young stellar population,
a highly asymmetric double-lobed FRII radio structure is possibly an archetypal example
to illustrate this process and to be used to investigate further the relationship between nuclear
and starburst activity.

\subsection{Comparison with other sources}

3C459 is clearly one of the most asymmetric sources with the ratio of separations of 
the outer hotspots,$r_{\rm D}$ defined to be $>$1, being $\sim$5 and the corresponding 
flux density ratio,$r_{\rm S}$, of the oppositely-directed lobes being $\sim$0.45 and 
0.3 at 5 and 1.7 GHz
respectively. From the compilation of symmetry parameters of high-luminosity 3CR and S4
sources (Saikia et al. 2001), there are only two objects with a separation ratio
$>$4, namely 3C254 and the compact steep-spectrum source B0428+205. 3C254 is associated
with a quasar at a redshift of 0.734 with $r_{\rm D}$=6.95 and $r_{\rm S}$=0.77 (Owen \& Puschell 1984;
Thomasson et al., in preparation).
Optical line and continuum imaging of this source shows an extended
emission-line region with the lobe on the nearer side 
interacting with a cloud of gas (Bremer 1997; Crawford \& Vanderriest 1997). Although
the flux density of the nearer lobe is brighter, the ratio is modest, which could
be a consequence of relativistic beaming of the hotspot further from the nucleus.
The CSS object B0428+205 which has a largest angular size of only 250 mas, is associated
with a galaxy at a redhsift of 0.219 with $r_{\rm D}$=4.69 and $r_{\rm S}$=0.16 (Dallacasa et al. 1995), 
consistent with a high dissipation of energy on the side where 
the jet interacts with a dense cloud.
Another example of a source with such a high degree of positional asymmetry is the
quasar 3C2 at a redshift of 1.037 for which $r_{\rm D}$=4.7 and $r_{\rm S}\sim$0.25 (Saikia, Salter
\& Muxlow 1987).  There is a sign of a radio jet extending from the core to
the northern lobe. Depolarization gradients in both the lobes suggest interaction 
with the external medium, but the northern lobe seems to be more significantly affected.

In the compilation of McCarthy, van Breugel \& Kapahi (1991), there are only four galaxies
with $r_{\rm D}>$ 4, namely, 3C99, 3C208.1, 3C459 and 3C460. All these sources exhibit a large
asymmetry in their flux density ratio, with the brighter lobe being closer to the nucleus.
In the case of 3C460, where they detect significant extended emission line gas, 
the surface brightness of this gas is much higher on the side of the lobe closer to the 
nucleus.  The radio galaxy 3C99 is similar to 3C459 in that it also has a steep-spectrum radio
core which was resolved by the EVN and shows multiple components. Its value of $r_{\rm D}$=4.8
while $r_{\rm S}\sim$0.04 and 0.03 at 5 and 1.7 GHz respectively, is consistent with the possibility that the
jet on the side of the nearer lobe is interacting with denser gas. Optical spectroscopic
observations along the axis of the source does indeed show that the gas on the side of the
nearer component is blue-shifted while that on the opposite side is red-shifted relative
to the galaxy. The blue-shifted gas is possibly approaching us, shifted outwards by
interaction with the radio jet (Mantovani et al. 1990).

\begin{figure}
\begin{center}
\vbox{
               \psfig{file=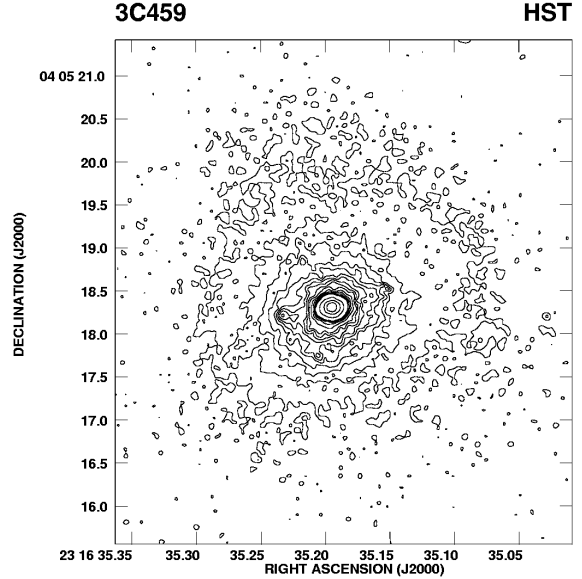,width=3.4in}
}
\end{center}
\caption{The HST image of 3C459 using the F702W filter. The contour levels
are 2.799$\times$10$^{-18}$$\times$(1, 2, 3, 4, 5, 6, 7, 8, 9, 10, 12, 14, 16, 18,
20, 40, 80, 160, 320, 640) erg cm$^{-2}$ s$^{-1}$ \AA$^{-1}$, while the peak value is
8.327$\times$10$^{-16}$ erg cm$^{-2}$ s$^{-1}$ \AA$^{-1}$.
}
\end{figure}

\begin{figure}
\hspace{1.0cm}
\begin{center}
\vbox{
               \psfig{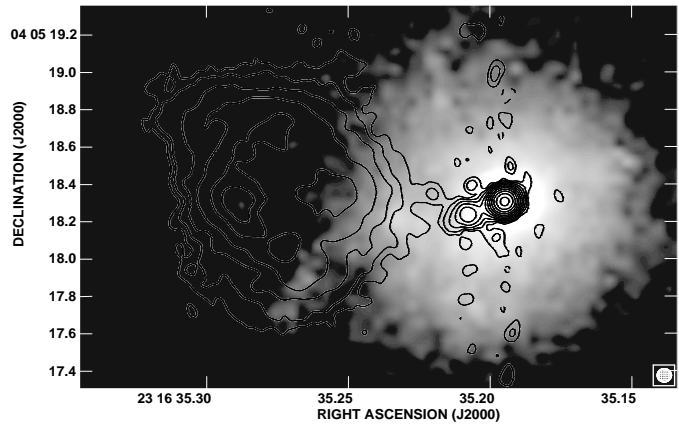}
}
\end{center}
\caption{The MERLIN image of the central and eastern components at 4866 MHz
with an angular resolution of 70 mas superimposed on the HST image show in 
grey scale. The contour levels for the radio image are the same as in Figure 3. 
}
\end{figure}

\section{Concluding remarks}
The radio galaxy, 3C459, with its young stellar population and high infrared luminosity 
may have undergone a recent starburst, possibly triggered by the merging of a companion galaxy.
The HST image presented here shows evidence of filamentary structures, which are possibly
of tidal origin, and a prominent peak of emission close to that of the nucleus 
of this N-galaxy, which could be due to galaxy in the late stages of merging. The
multifrequency radio observations using MERLIN, VLA and the EVN, clarify the small- and
large-scale structure of the source, which is highly asymmetric. The eastern component, 
which is brighter, closer to
the nucleus and more strongly depolarized is interacting with denser gas, possibly 
related to the merging process. This lobe has a shell-like structure with a prominent
hotspot and two secondary peaks of emission. The MERLIN and EVN observations show the
central component to consist of several sub-components whose orientations are suggestive
of a helical or strongly bent jet. However, we have not been able to identify a flat-spectrum
radio core from these observations. It may be possible to do this with multifrequency, mas-resolution
observations. 3C459 is an archetypal example of a high-luminosity, compact steep-spectrum
radio galaxy exhibiting evidence of a starburst. Identification of a larger sample of such
objects could enhance our understanding of the relationship between these two forms
of activity.

\section*{Acknowledgments}
We thank Jim Ulvestad for giving us the calibrated VLA A-array data at 5 GHz, Neal Jackson
for help in analysing the HST data, Martin Krause for useful discussions, 
Anita Richards for help in producing one of the figures, an anonymous referee for drawing
our attention to some relevant references,
and the staffs of the Cambridge One-Mile Telescope, MERLIN, Effelsberg, Medicina, 
Noto, Onsala and Westerbork for their assistance during the course of these observations.
One of us (DJS) would like to thank the PPARC Visitors Programme at Jodrell Bank Observatory
and Ralph Spencer who looks after this programme for financial support, Andrew Lyne, Director,
for use of the facilities at the Observatory, and Peter Thomasson for hospitality while this
work was done.  MERLIN is a U.K. National Facility operated by the University of Manchester 
on behalf of PPARC.  The Very Large Array is operated by the National Radio Astronomy
Observatory for Associated Universities Inc. under a co-operative
agreement with the National Science Foundation.  This research used observations with
the Hubble Space Telescope, obtained at the Space Telescope Science Institute, which is operated 
by Associated Universities for Research in Astronomy, Inc., under NASA contract NAS 5-26555. 
This research has made use of the NASA/IPAC extragalactic database (NED)
which is operated by the Jet Propulsion Laboratory, Caltech, under contract
with the National Aeronautics and Space Administration.

{}

\end{document}